\documentclass[review]{elsarticle}

\usepackage{lineno,hyperref}
\modulolinenumbers[5]
\usepackage{amstext}
\usepackage{amssymb}   
\usepackage{amsmath}   
\journal{Physics Letters B}

\begin{document}

\begin{frontmatter}

\title{Gravity and nonabelian gauge fields in noncommutative space-time}

\author[mymainaddress,mysecondaryaddress]{Nguyen Ai Viet\footnote{Corresponding author, E-mail:naviet@vnu.edu.vn}} \author[mysecondaryaddress]{Pham Tien Du\footnote{E-mail: doandu30111989@gmail.com,}}  



\address[mymainaddress]{Information Technology Institute, Hanoi National University, 144 Xuan Thuy Boulevard, Cau Giay, Hanoi, Vietnam}
\address[mysecondaryaddress]{Department of Physics, College of Natural Sciences, Hanoi National University, 334 Nguyen Trai, Thanh Xuan, Hanoi, Vietnam}

\begin{abstract}
Noncommutative geometric construction of gravity in the two sheeted spacetime can be viewed as a discretized version of a Kaluza-Klein theory. In this paper, we show that it is possible to incorporate the nonabelian gauge fields in the same framework. The generalized Hilbert-Einstein action is gauge invariant only in two cases. In the first case, the gauge group must be abelian on one sheet of spacetime and nonabelian on the other one. In the second case, the gauge group must be the same on two sheets of spacetime. Actually, the theories of electroweak and strong interactions are exactly these two cases.
\end{abstract}


\end{frontmatter}

Noncommutative geometry (NCG) was proposed by Connes \cite{Connes} as a new concept of spacetime. In particular, NCG can treat the discrete dimensions on equal footing with the continuous ones. In spite of its initial success in applying this idea in the Standard Model \cite{CoLo}, the progress in constructing the Einstein theory of gravity based on NCG has not been as straightforward. Gravity coupled to Yang-Mills gauge and Higgs fields can be derived from the spectral action \cite{NSA} in a perturbative expansion, however the full physical content of the theory are not transparent form the first principles, it must be introduced by hand in the Dirac operator. In the most recent reviews on the matter \cite{Review2012}, while the technical aspects are developed impressively, the physical contents have not been addressed at the same level. Hence, the idea of NCG has not been widely known in the recent developments of modified gravity to address the cosmological problems \cite{DeRham}.

In 1993, Chamsedine, Felder and Fr\"ohlich \cite{CFF} have made the first attempt in constructing gravity in noncommutative space-space. Specialized in the ordinary space-time extended by a discrete space-like extra dimension consisting of two points, which was original proposed by Connes and Lott \cite{CoLo}, this construction has resulted in a rather disappointing no-go theorem, stating that the most general NCG in the Connes-Lott's spacetime leads to no new physical content.

In 1994, Landi, Viet and Wali \cite{LVW}, have overcome the no-go theorem and succeeded in constructing a generalized Einstein-Hilbert action in Connes-Lott's spacetime leading to the zero mode spectrum of Kaluza-Klein theory \cite{KK}. This result has inspired Viet and Wali to utilize the idea of using NCG as a discretized version of Kaluza-Klein theory \cite{VW}. In the most general form of this theory the gravitational, vector and scalar fields emerge in pairs. In each pair, one field is massless and the other one is massive. Chamsedine-Felder-Fr\"ohlich's \cite{CFF} and Landi-Viet-Wali's \cite{LVW} models are just special cases of this general framework. The advantage of the discretized Kaluza-Klein theory over the usual one is that it does not contain finite towers of massive fields, which can lead to both theoretical and observational inconsistencies.

In 1968, R.Kerner \cite{Kerner} has shown that nonabelian gauge fields can be incorporated into the Kaluza-Klein theory as components of gravity. Landi-Viet-Wali's model cannot be generalized to the nonabelian case, because it does not contain the quartic terms of the vector fields.

In this paper, we show that it is possible to generalize Viet-Wali's construction with minor modifications to include covariant action of nonabelian gauge fields as parts of the generalized Hilbert-Einstein action. It is interesting that the gauge invariance can be kept in some specific choices of gauge groups. Actually, the gauge theories of electroweak and strong interactions can fit into these cases. Moreover, the gauge fields are then all massless. They will acquire masses when Higgs mechanism is turned on leading to a unified theory of all the known interactions and Higgs fields in the recent paper \cite{QuyNhon}.

Let us follow the steps of \cite{VW} to construct the model. The spectral triplet of NCG is given as follows
 
i) A Hilbert space which is sum of two Hilbert spaces $ {\cal H} = { \cal H}_L \oplus {\cal H}_R$. Thus, the spinors in our model are direct sums of two spinors of the 4-dimension and represented as follows 
\begin{equation}
\Psi =
\begin{bmatrix}
    \psi_L \\
    \psi_R
\end{bmatrix},
\end{equation}
where the sub-indexes $I=L,R$ are used to denote the the chiral spinors.

ii) The algebra $ {\cal A} = {\cal A}_L \oplus {\cal A}_R$, where ${\cal A}_I= {\cal C}^\infty({\cal M}^4)$. So, the elements of ${\cal A}$ can be represented as the diagonal 0-form matrices ${\cal F}$ as follows
\begin{equation}\label{Function}
F(x) = 
\begin{bmatrix}
    f_L(x) & 0 \\
    0 & f_R(x)
\end{bmatrix},
\end{equation} 
where $f_I$ are real functions, which are elements of ${\cal A}_I$. 
The 0-forms can also be represented as elements of $Z_2$ algebra as follows
\begin{equation}\label{Function2}
F(x) = f_+(x) {\bf e} + f_-(x) {\bf r} ~,~ f_\pm(x) = {1\over 2}(f_L(x) \pm f_R(x)),
\end{equation} 
where
\begin{eqnarray}
{\bf e}= \begin{bmatrix}
1 & 0 \cr
0 & 1 
\end{bmatrix} ~,~ {\bf r}= \begin{bmatrix}
1 & ~0 \cr
0 & -1 
\end{bmatrix} \nonumber \\
{\bf e}^2 ={\bf r}^2 = {\bf e}~,~ {\bf e}{\bf r} = {\bf r}{\bf e}={\bf r}
\end{eqnarray} 
iii) A self-adjoint Dirac operator $D = d. {\bf e} + \Theta $, where $d= \gamma^\mu \partial_\mu$ is the usual Dirac operator and $D^2 = d \Theta + \Theta d = \Theta^2 = 0 $. Thus, $D$ is represented on the Hilbert space ${\cal A}$ as follows
\begin{eqnarray}
& D &= DX^M D_M = DX^\mu \partial_\mu + DX^5 \sigma^\dagger D_5\nonumber \\
& DX^\mu & =
\begin{bmatrix}
\gamma^\mu & 0\\
0 &  \gamma^\mu
\end{bmatrix} ~,~
 DX^5 =
\begin{bmatrix}
0 & i\gamma^5 \\
-i \gamma^5 & 0
\end{bmatrix},
\end{eqnarray}
where $\mu = 0,1,2,3 ; M= \mu, 5$, $\gamma^\mu$ and $\gamma^5$ are the usual Diract matrices, which represent differential elements of ${\cal M}^4$ in the above representation. 
The action of $D_5$ on the 0-forms ${\cal F}$ is defined as follows
\begin{equation}\label{DER5}
D_5 F(x) = m (f_L(x)-f_R(x)){\bf r}= 2m f_-(x){\bf r}, 
\end{equation}
where $m$ is a mass dimensioned parameter. The derivation defined in Eq.(\ref{DER5}) satisfies the Leibnitz rule
\begin{equation}
D_5 (FG) = D_5F.G + F.D_5G 
\end{equation}

The action of the Dirac operator $D$ on the 0-forms $F$ give the closed 1-forms $DF$ as follows
\begin{equation}\label{eq:closed}
DF(x) = 
\begin{bmatrix}
df_1(x)   &  -2 i m \gamma^5 f_-(x)\\
2 i m \gamma^5 f_-(x)  & df_2(x) 
\end{bmatrix}
\end{equation}

The algebra of 1-forms of this NCG must be defined as an extension of the set of closed forms in Eq.(\ref{eq:closed}) to the right module over ${\cal A}$ as follows
\begin{equation}\label{eq:1form}
U(x) = 
\begin{bmatrix}
\gamma^\mu u_{\mu L}(x) & \gamma^5 u_{5L}(x) \\
\gamma^5 u_{5R}(x) & \gamma^\mu u_{\mu R}(x) 
\end{bmatrix} = DX^\mu U_\mu + DX^5 U_5
\end{equation}
where $U_\mu$ and $U_5$ are real 0-forms defined in Eq.(\ref{Function}).

To define the 2-forms, one have to define the wedge product of two 1-forms to satisfy the following anti-commutativity condition as in the ordinary differential geometry
\begin{eqnarray}\label{wedge1}
U \wedge V &=& - V \wedge U \nonumber \\
DX^M \wedge DX^N &=& - DX^N \wedge DX^M
\end{eqnarray}
The general 2-forms are defined as the right module over the algebra of 0-forms
\begin{eqnarray}
S &=& DX^M \wedge DX^N S_{MN},  
\end{eqnarray}
where the components $S_{MN}$ are 0-forms.
The wedge product of two 1-forms can be defined in its components as follows
\begin{eqnarray}
(U \wedge V)_{\mu \nu} &=& - (U \wedge V)_{\nu \mu} = {1 \over 2}(U_\mu V_\nu - U_\nu V_\mu) \nonumber \\
(U \wedge V)_{\mu 5} &=& - (U \wedge V)_{5 \mu} = {1 \over 2}({\tilde U}_\mu V_5 - U_5 V_\mu), 
\end{eqnarray}
where the "\~{}" operation is defined as
\begin{equation}
{\tilde F} = f_+ {\bf e} - f_-{\bf r}
\end{equation}
The action of the Dirac operator $D$ on a 1-form $U$ will give a 2-form $DU$ with the following components
\begin{eqnarray} \label{1formder}
(DU)_{\mu \nu} &=& {1 \over 2}(\partial_\mu U_\nu - \partial_{\nu} U_\mu) \nonumber \\
(DU)_{\mu 5} &=& {1 \over 2}(\partial_\mu U_5 - m U_{-\mu}) = (DU)_{5\mu}
\end{eqnarray}
Eq.(\ref{1formder}) is consistent with the De Rham's condition $D^2=0$.

The generalized Riemannian geometry is constructed in the following steps:

1.{\it Step 1}: The general curved basis $DX^M$ related to the
local orthonormal basis $E^A(x), A = a, \dot{5}; a =0,1,2,3$ by the invertible vielbein $E^A_M(x)$ with the following transformations
\begin{eqnarray}\label{eq:transform}
E^A = DX^M E^A_M(x) &,&
DX^M = E^A E^M_A(x) \nonumber \\
E^A_M(x) E^N_A(x) = \delta^N_M(x) &,& E^A_N(x) E^N_B(x) = \delta^A_B(x) 
\end{eqnarray}
The inner scalar product of 1-forms ${\cal U}$ and ${\cal V}$ is defined as follows
\begin{eqnarray}\label{metric}
&<DX^M, DX^N> = G^{MN}(x)& \nonumber \\
&<U, V> =  U^\dagger_M G^{MN}(x) U_N&
\end{eqnarray}
It is always possible to choose the vielbein coefficients $E^A_M(x)$ so that the basis $E^A$ satisfies the following locally flat condition
\begin{equation}\label{eq:flatcond}
<E^A,E^B> = \eta^{AB} = diag(-1,1,1,1,1)
\end{equation}
Applying Eq.(\ref{metric}), we can express the locally flat condition (\ref{eq:flatcond}) as an expression of metric in terms of vielbein as follows
\begin{equation}\label{eq:metric}
G^{MN}(x) = E^M_a(x) \eta^{ab} E^N_b(x) + E^M_{\dot{5}}(x) E^N_{\dot{5}}(x)
\end{equation}
2.{\it Step 2}: Without any loss of generality, one can generalize Viet-Wali's vielbein as follows
\begin{eqnarray}\label{eq:genvielbein}
E^a_\mu(x) = \begin{bmatrix}
e^a_{\mu L}(x)& 0 \cr
0 & e^a_{\mu R}(x)
\end{bmatrix}
, E^{\dot{5}}_{5} = 
\begin{bmatrix}
\phi_{L}(x)& 0 \cr
0 & \phi_{R}(x) 
\end{bmatrix},&&  \cr
E^{\dot{5}}_\mu(x) = 
-\begin{bmatrix}
a_{\mu L}(x)\phi_L(x)& 0 \cr
0 & a_{\mu R}(x)\phi_R(x) 
\end{bmatrix},E^a_5 = 0, &&
\end{eqnarray} 
where $a_{I\mu}(x)= a^i_{I\mu} T^i_I, i=1,..., n_I$ are nonabelian gauge fields and $T^i_I$ are generators of the gauge group $G_I$.

The inverse vielbein matrix elements are derived from Eq.(\ref{eq:transform}) as follows
\begin{eqnarray}\label{eq:inversevielbein}
 E^\mu_a(x) = \begin{bmatrix}
e^\mu_{a L}(x)& 0 \cr
0 & e^\mu_{a R}(x)
\end{bmatrix}
,& E^\mu_{\dot 5}(x) = 0,&  \nonumber \\
 E^5_a(x) = - E^\mu_a(x) A_\mu(x) \Phi^{-1}(x)
,& E^5_{\dot 5} = \Phi^{-1}(x)
& 
\end{eqnarray} 
The metric and its inverse can be calculated from Eqs.{\ref{eq:genvielbein}) and {\ref{eq:inversevielbein}) using the following relations
\begin{eqnarray}\label{METRIC}
{\cal G}^{MN} &~=~& E^M_A\eta^{AB}E^N_B \\
{\cal G}_{MN} &~=~& E^A_M \eta_{AB}E^B_N
\end{eqnarray}

3.{\it Step 3}: The Levi-Civita connection 1-forms $\Omega^A_{~B}$ are defined with the covariant derivative
\begin{equation}
\nabla E^A =E^B \otimes \Omega^A_{~B},
\end{equation}
The Levi-Civita connection is called metric compatible if it satisfies the condition
\begin{equation}\label{metriccomp}
\Omega^\dagger_{AB} = - \Omega_{BA}
\end{equation}
The metric compatibility condition (\ref{metriccomp}) leads to the following restriction on the components
\begin{eqnarray}\label{comp2}
\Omega_{~abc} & ~=~ & -~\Omega^{bac}, \cr
\Omega_{~ab {\dot 5}} & ~=~ & -~\Omega_{~ba{\dot 5}}~ = ~
\omega_{~ab {\dot 5}} e~~,\cr
\Omega_{~a{\dot 5}b} & ~=~ & - ~\Omega_{~{\dot 5}ab}, \cr
\Omega_{~a{\dot 5}{\dot 5}} & ~=~ & -~ \Omega_{~{\dot 5}a{\dot 5}} ~=~
\omega_{~a{\dot 5}{\dot 5}} e,\cr
\Omega_{~{\dot 5}{\dot 5}a}&~ =~ & \Omega_{~{\dot 5}{\dot 5}{\dot 5}} ~=~0 ~~.
\end{eqnarray}
 
4.{\it Step 4}: The Cartan structure equations can be generalized as follows
\begin{eqnarray}
T^A &=& DE^A - E^B \wedge \Omega^A_{~B} \label{eq:struct1} \\
R^{AB}&=& D\Omega^{AB} + \Omega^A_{~C} \wedge \Omega^{CB} \label{eq:struct2}
\end{eqnarray}
It is shown in Ref.\cite{VW, Constraint} in order to determine Levi-Civita connection and the torsion in terms of vielbein coefficients from the the first structure equation (\ref{eq:struct1}) and the metric compatibility condition (\ref{comp2}), one must impose the following constraints 
\begin{eqnarray}\label{TORFREE}
T_{abc}=T_{ab{\dot 5}}=0 &~,~& T_{{\dot 5}AB}= t_{{\dot 5}AB} r .
\end{eqnarray}
From now on we will specialize with the ansatz
\begin{eqnarray}\label{ansatz}
&& E^a_\mu(x)= e^a_\mu(x) {\bf 1}~,~ \Phi(x) = \phi(x) {\bf 1}, \nonumber \\
&& A_\mu(x)= a_{+\mu} {\bf 1}+ a_{-\mu} {\bf r} = 
\begin{bmatrix}
a_{\mu L}(x) & 0 \cr
0 & a_{\mu R}(x)
\end{bmatrix} 
\end{eqnarray}
The Levi-Civita connection 1-forms can be calculated explicitly as follows
\begin{eqnarray}
\Omega_{abc} 
&=&  \omega_{abc} \\
\Omega_{ab\dot 5}
&=&
{1 \over 4}e^\mu_{~a} e^\nu_{~b}\phi(F_{\mu \nu} +{\tilde F}_{\mu \nu} + m [{\tilde A}_\mu - A_\mu, {\tilde A}_\nu- A_\nu]) \\
\Omega_{\dot 5bc}
& = & -{1 \over 4} e^\mu_{~b} e^\nu_{~c}\phi(F_{\mu \nu} + {\tilde F}_{\mu \nu}
+ m [{\tilde A}_\mu - A_\mu, {\tilde A}_\nu- A_\nu]) \\ 
\Omega_{\dot 5 b \dot 5} 
&=& 
 e^\mu_{~b}
{\partial_\mu \phi \over\phi}
\end{eqnarray}
Then it is straightforward to calculate the curvature tensor in terms of the Levi-Civita connections using the second structure equation (\ref{eq:struct2}).

5.{\it Step 5}: The generalized Ricci scalar curvature is given as follows
\begin{equation}
R = Tr(\eta^{AC}\eta^{BD}R_{ABCD}),  
\end{equation}
where the trace is taken on both $Z_2$ and nonabelian group indices.
The generalized Hilbert-Einstein's action is defined as follows
\begin{eqnarray}\label{HEaction}
S &=& {1 \over 16 \pi G_N}\int dx^4 \sqrt{-det|g|} \phi (x) Tr(R),
\end{eqnarray}
where $G_N$ is the Newton constant.

It is straight forward to calculate the Ricci scalar curvature in terms of gravity, nonabelian gauge fields and Brans-Dicke scalar $\phi$ as follows
\begin{eqnarray}\label{5Ricci}
R  &=& R_4 - {{2 \Box \phi} \over \phi} + {\cal L}_g  \\
{\cal L}_g &=& - {1 \over 4} \phi^2 g^{\mu\nu} g^{\rho\tau} \hat{f}_{\mu\nu} \hat{f}_{\rho \tau}, 
\end{eqnarray}
where $\Box = g^{\mu \nu}(x) \nabla_\mu \partial_\nu$ and $\nabla_\mu$ is the covariant derivative and
\begin{equation}
\hat{f}_{\mu\nu} = \partial_\mu a_{+\nu}- \partial_\nu a_{+\mu} 
+ 2m [a_{-\mu}, a_{-\nu}]\label{gstrength} ,
\end{equation}

The Ricci scalar curvature in Eq.(\ref{5Ricci}) is gauge invariant only  in the following cases:

{\bf a. Case 1}: 
\begin{eqnarray}\label{weak}
&& a_{\mu R} = {1 \over M_2} B_\mu(x) ~,~ a_{\mu L} = {1 \over M_1} (B_\mu(x)- {M_1 g \over m} W_\mu(x)) \nonumber\\
&& a_{-\mu} = {M_2-M_1\over 2M_1 M_2} B_\mu(x) - {g \over 2m} W_\mu(x) \nonumber \\
&& a_{+\mu} = {M_2+M_1\over 2M_1 M_2} B_\mu(x) - {g \over 2m} W_\mu(x)
\end{eqnarray}
where $B_\mu(x)$ is an abelian gauge vector and $W_\mu(x) = W^i_\mu(x) T^i$ are nonabelian gauge vectors, $M_1$ and $M_2$ are two mass parameters.  
\begin{eqnarray}
\hat{f} & = & {M_2+M_1\over 2M_1 M_2} F_{\mu \nu} + {g \over 2m} G_{\mu \nu}\cr
F_{\mu\nu}&=& \partial_\mu B_\nu - \partial_\nu B_\mu \cr
G_{\mu\nu} &=& \partial_\mu W_\nu - \partial_\nu W_\mu +g[W_\mu, W_\nu] 
\end{eqnarray}
The gauge part ${\cal L}_g$ in the Hilbert-Einstein action in Eq.(\ref{HEaction}) now can be calculated in terms of the physical field strengths as follows
\begin{eqnarray} \label{SGWEAK}
&& S_g= \int dx^4 \sqrt{-det g} - {1 \over 4} (F^2 + G^2) 
 = {1 \over 16 \pi G_N} \cr
&&\int dx^4  
 \sqrt{-det g}
 - {1 \over 4} ({(M_2+M_1)^2 \over 4M^2_1 M^2_2} F^2 
 + {g^2 \over 4 m^2} G^2),
\end{eqnarray}
Hence we have the following relation between the parameters
\begin{eqnarray}
&& g = 8 m \sqrt {\pi G_N} ~,~ {M_2+M_1 \over M_1 M_2} = 8 \sqrt{\pi G_N}
\end{eqnarray}
The gauge action $(\ref{SGWEAK})$ is gauge invariant. 

Let us specialize with the abelian gauge group $U(1)$ and the nonabelian one $SU(2)$. The coupling of the gauge fields to the chiral spinor matter fields is given in the following action \cite{VW2000} 
\begin{equation}
S_{g-\Psi} = \int dx^4 \sqrt{-det|g|} <\Psi|\gamma^a e^\mu_a(x) iA_\mu |\Psi>
\end{equation}
which reduces to the following action of the electroweak gauge fields coupled to the chiral quarks and leptons
\begin{eqnarray}
&& S_{g-\Psi}= \int dx^4 \sqrt{-det|g|} (<\psi'_R|\gamma^a e^\mu_a(x) i g' B_\mu |\psi'_R> \cr
&& + {i \over 2} <\psi'^\alpha_L|\gamma^a e^\mu_a(x) g' B_\mu \delta^\beta_\alpha - g W^i_\mu(x)(\sigma^i)^\beta_\alpha |\psi'_{L\beta}>,  
\end{eqnarray}
if we define the chiral spinor fields as follows
\begin{eqnarray}
\psi'_L = \sqrt{2 \over g'M_1}\psi_R &~,~& 
\psi'_R = {1 \over \sqrt{g'M_2}}\psi_R
\end{eqnarray}
and the following relation between $m$ and $M_1$ must hold
\begin{equation}
m = 2M_1
\end{equation}

{\bf b. Case 2}: 
Let us consider the following physical nonabelian gauge field $C_\mu(x) = C^i_\mu(x) \lambda^i$, which are introduced via following relations
\begin{eqnarray}
a_{\mu L}(x)&=& {1 \over M_3} C_{\mu}(x)~,~ a_{\mu R}(x)= {1 \over M_4} C_{\mu}(x) \\
 a_{\pm\mu}(x)&=& {M_4 \pm M3 \over 2M_3 M_4} C_{\mu}(x)
\end{eqnarray}
Hence, we have
\begin{eqnarray}
\hat{f}_{\mu \nu}&=& {M_4+M_3 \over 2M_3 M_4} H_{\mu \nu} \cr
H_{\mu \nu}&=& \partial_\mu C_\nu(x) - \partial_\nu C_\mu(x)
+ g [C_\mu(x), C_\nu(x)]
\end{eqnarray}
where the nonabelian gauge coupling constant $g$ is 
\begin{equation}
g = {m (M_4-M_3)^2 \over (M_4+M_3)M_3 M_4} 
\end{equation}
The gauge action in the generalized Hilbert-Einstein action is
\begin{equation} \label{SGSTRONG}
S_g= \int dx^4 \sqrt{-det|g|} - {1 \over 4} H^2
\end{equation}
if the following relation holds
\begin{equation}
{M_4+M_3 \over 2M_3 M_4} = 8 \sqrt{\pi G}
\end{equation}
The action term coupling the gauge and spinor fields is given as follows
\begin{eqnarray}\label{Smatter}
S_{g-\Psi}&= & ig \int dx^4 \sqrt{-det|g|} Tr<\Psi| \gamma^a e^\mu_a(x) A_\mu(x)|\Psi> \cr
&=& ig \int dx^4 \sqrt{-det|g|} (<\psi'^\alpha_L|C^i_\mu(x)(\lambda^i)^\beta_{\alpha}|\psi'_{L\beta}> \cr 
&& + <\psi'^\alpha_R|C^i_\mu(x) (\lambda^i)^\beta_{\alpha}|\psi'_{R\beta}> ),
\end{eqnarray}
where $(\lambda^i)^\beta_\alpha$ matrices are a representation of the nonabelian gauge group's generators and we have redefined the spinors with the physical ones as follows
\begin{eqnarray}
\psi'_L = {1 \over \sqrt{M_3}} \psi_L &~,~&
\psi'_R = {1 \over \sqrt{M_4}} \psi_R. 
\end{eqnarray}
If we choose the color $SU(3)$ as the nonabelian gauge groups, then the action (\ref{SGSTRONG}) becomes the action of QCD \cite{ChengLi} coupled to gravity. The chiral spinors now can be choosen as quarks with the color indices $\alpha,\beta=r,b,y$. The action $S_{g-\Psi}$ has a chiral symmetry $SU(3)XSU(3)$ as QCD does in the chiral limit. 

Let us summarize the results and discuss the physical content of our model. In this paper, we have successfully constructed the generalized Hilbert-Einstein action of two sheeted noncommutative spacetime ${\cal M}^4 \times Z_2$. The action can contain covariant terms for nonabelian gauge fields in only two cases. In the first case, the gauge group must be abelian on one sheet, which can be applied for the gauge theory of electroweak interaction. In the second case, the gauge groups must be the same in both sheets, which can be applied to the QCD. In some sense, NCG can explain why the strong interaction has the chiral symmetry $SU(3) \times SU(3)$ and the nonabelian $SU(2$ gauge fields do not couple to the right handed chiral spinor fields.

Our model also implies some relations between the Newton constant, weak coupling constant and the mass dimensional parameters of the model, which are very similar to the results obtained in some previous works on different context \cite{MRST94}. The physical interpretation of these results will merit further exploration along this direction. The first result is a unified theory of all known interactions and Higgs fields as components of gravity in noncommutative spacetime \cite{QuyNhon}.

Thanks are due to Nguyen Van Dat (ITI-VNU) and Do Van Thanh (Department of Physics, College of Natural Sciences, VNU) for their discussions. The supports of ITI, VNU and Department of Physics, College of Natural Sciences, VNU are appreciated.

\end{document}